\documentclass[letter]{jpsj3}
\usepackage{color}

\title{%
Pressure-induced Enhancement of Superconductivity in BiS$_2$-layered LaO$_{1-x}$F$_x$BiS$_2$
}

\author{%
Takahiro TOMITA$^{1}$\thanks{E-mail address: tomita@phys.chs.nihon-u.ac.jp:Ver. 14},
Masaya EBATA$^{1}$,
Hideto SOEDA$^{1}$,
Hiroki TAKAHASHI$^{1}$,
Hiroshi FUJIHISA $^{2}$,
Yoshito GOTOH$^{2}$,
Yoshikazu MIZUGUCHI$^{3}$,
Hiroki IZAWA$^{3}$,
Osuke MIURA$^{3}$,
Satoshi DEMURA$^{4}$, 
Keita DEGUCHI$^{4}$, 
Yoshihiko TAKANO$^{4}$
}

\inst{%
$^{1}$Department of Physics, College of Humanities and Sciences, Nihon University, Sakurajousui, Setagaya-ku, Tokyo 156-8550, Japan\\
$^{2}$National Institute of Advanced Industrial Science and Technology (AIST), Tsukuba Central 5, 1-1-1, Higashi, Tsukuba 305-8565, Japan\\
$^{3}$Department of Electrical and Electronic Engineering, Tokyo Metropolitan University, 1-1, Minami-osawa, Hachioji, 192-0397, Japan\\
$^{4}$National Institute for Materials Science, 1-2-1, Sengen, Tsukuba 305-0047, Japan\\
}

\recdate{\today}

\abst{%
The newly discovered BiS$_2$-based LaO$_{1-x}$F$_{x}$BiS$_2$ ($x$=0.5) becomes superconductive at $T_c$=2.5 K.
Electrical resistivity and magnetization measurements are performed under pressure to determine the pressure dependence of the superconducting transition temperature $T_c$.
We observe that $T_c$ abruptly increases from 2.5 K to 10.7 K at a pressure of 0.7 GPa.
According to high-pressure X-ray diffraction measurements, a structural phase transition from a tetragonal phase ($P$4/$nmm$) to a monoclinic phase ($P$2$_1/m$) also occurs at around $\sim$ 1 GPa.
We consider that a pressure-induced enhancement of superconductivity is caused by the structural phase transition.  
}

\kword{%
superconductivity, oxypnictides, pressure, magnetization, transport
}

\begin{document}
\maketitle

\newpage
\section{\label{sec:level1}INTRODUCTION}
A wide variety of layered materials have been found to be superconducting, such as cuprates with CuO$_2$ layers and iron pnictide superconductors with FeAs layers. Both layered superconductors have gained much attention because each has a high superconducting transition ($T_c$), beyond 40 K.
The newly discovered BiS$_2$-based layered materials Bi$_4$O$_4$S$_3$ and $Re$O$_{1-x}$F$_{x}$BiS$_2$ ($Re$=La, Ce, Pr, and Nd) is superconducting at low temperatures \textcolor{blue}{\cite{Mizuguchi2012-4, Mizuguchi2012-2, Demura, Hechang, Li2012, awana, Mizuguchi2012}}.
The crystal structure of Bi$_4$O$_4$S$_3$ is a stack of Bi$_4$O$_4$(SO$_4$) and Bi$_2$S$_4$ layers (double BiS$_2$ superconducting layers) and can be obtained at $T_c$=4.5 K.
On the other hand, the LaO$_{1-x}$F$_{x}$BiS$_2$ superconductor has a structure of stacked BiS$_2$ superconducting and LaO$_{1-x}$F$_x$ block layers, which resembles the layered structures of high-$T_c$ cuprate superconductors and the iron pnictide superconductors. Interestingly, the space group ($P4/nmm$) for the crystal structure LaO$_{1-x}$F$_{x}$BiS$_2$ is the same as that of the layered iron-based superconductors LaO$_{1-x}$F$_x$FeAs \textcolor{blue}{\cite{kamihara}}.
The transition temperature $T_c$ for LaO$_{1-x}$F$_{x}$BiS$_2$ only appears with 50\% F substitution at the O sites, causing electron doping. 
The polycrystalline samples of LaO$_{0.5}$F$_{0.5}$BiS$_2$ were synthesized by two different methods, `AP' and `HP'. 
LaO$_{0.5}$F$_{0.5}$BiS$_2$ with $T_c$=2.5 K was synthesized by annealing the sample under ambient pressure (`AP'). On the other hand, LaO$_{0.5}$F$_{0.5}$BiS$_2$ with $T_c \sim$10.5 K was synthesized by first synthesizing BiS$_2$LaO$_{0.5}$F$_{0.5}$ (AP) and then performing additional annealing under high pressure (`HP'). In a recent study of high-pressure measurements of LaO$_{0.5}$F$_{0.5}$BiS$_2$ (HP), the $T_c$ was shown to increase from 9.6 K under ambient pressure to 10.5 K under 1 GPa, which then gradually decreased with application of pressure beyond 1 GPa \textcolor{blue}{\cite{kotegawa}}. 
It is strange that LaO$_{1-x}$F$_{x}$BiS$_2$ has been shown to have two different $T_c$ values with the same $x$ value.
Therefore, in this letter, we first applied pressure to LaO$_{0.5}$F$_{0.5}$BiS$_2$ (AP) at a low $T_c$(=2.5 K) and investigated the pressure dependence of $T_c$ and the crystal structure to compare two kind of materials, AP- and HP-processed LaO$_{0.5}$F$_{0.5}$BiS$_2$.

\section{\label{sec:level2}EXPERIMENTAL DETAILS}
A polycrystalline sample of BiS$_2$LaO$_{0.5}$F$_{0.5}$ (AP) was synthesized using a solid-state reaction at ambient pressure \textcolor{blue}{\cite{Mizuguchi2012-4, Mizuguchi2012-2}}. Electrical resistivity measurements under a varying hydrostatic pressure (pressure medium: Daphne$^{\mathrm{TM}}$ 7474) up to 3 GPa were performed using a standard four-probe method in a piston-cylinder-type pressure cell. Further external pressure up to 18 GPa was applied using a diamond anvil cell (DAC). In this case, the sample chamber was sealed with an Re gasket and filled with powdered NaCl as a pressure-transmitting medium.\textcolor{blue}{\cite{Hazen}}. DC magnetization measurements were carried out for BiS$_2$LaO$_{0.5}$F$_{0.5}$ (AP) samples up to 1 GPa using a SQUID magnetometer (Quantum Design, MPMS-XL7) with a piston-cylinder cell (pressure medium:  Daphne$^{\mathrm{TM}}$ 7474) \textcolor{blue}{\cite{uwatoko}}. The superconducting temperature of lead (2.6 mg) mounted with the sample (29 mg) was used as a manometer to check the pressure. 

To investigate the crystal structure of BiS$_2$LaO$_{0.5}$F$_{0.5}$ (AP), a Rigaku D/MAX 2000 X-ray diffractometer with a conventional MoK  rotating anode generator (Rigaku, Co. Ltd., normal focus, 40 kV/50 mA,  =0.7107 \AA) was used at room temperature, using the DAC with a pressure-generating membrane and a 60-$\mu$m-thick stainless-steel gasket. The powder sample was loaded into the gasket hole (diameter 0.3$\phi$) with Daphne$^{\mathrm{TM}}$ 7474 as the pressure-transmitting medium. The pressure was calibrated by the ruby luminescence method.
We also carried out an angle-dispersive powder x-ray diffraction experiment under pressure at beamline BL-18C at the Photon Factory at the Institute of Materials Structure Science (KEK) in Japan. The beams were monochromatized to a wavelength of 0.6181 \AA \ and collimated to a 100 $\mu$m diameter. Powder diffraction patterns were detected by an imaging plate detector with a typical exposure time of 15 min and a sample-detector distance of about 250.00 mm. Two-dimensional Debye--Scherrer rings were obtained and converted to one-dimensional 2-intensity patterns by using PIP software \cite{Fujihisa2009}.  

\section{\label{sec:level3}RESULTS AND DISCUSSION}
\begin{figure}[h!]
\begin{center}
\includegraphics[width=30pc]{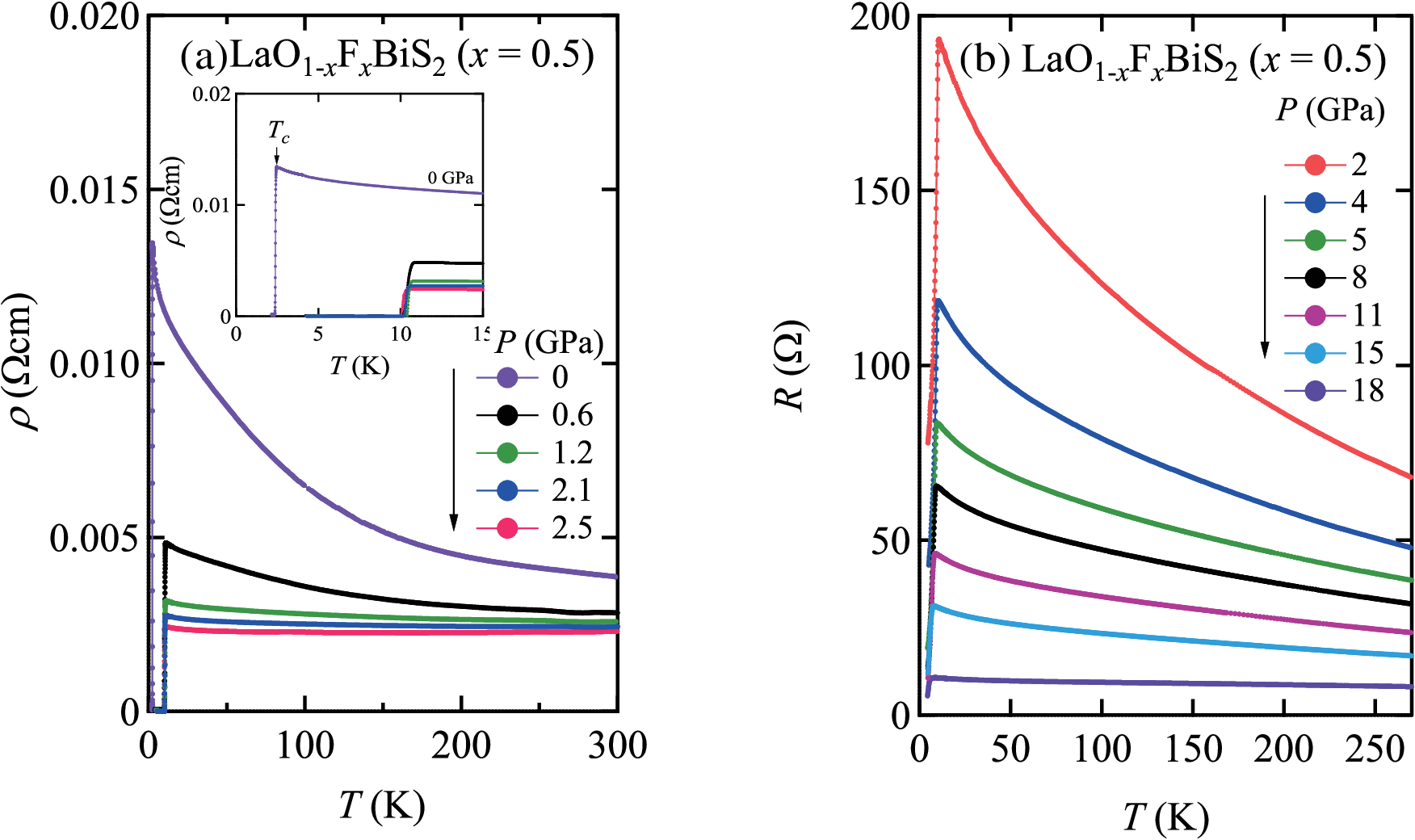}
\end{center}
\caption{(Color online) 
Pressure dependence of electrical resistivity $\rho (T)$ for LaO$_{0.5}$F$_{0.5}$BiS$_2$  using (a) a piston-cylinder cell up to 3 GPa and 
(b) a diamond anvil pressure cell up to 18 GPa. A direct current of 1 mA is applied. The arrow in the inset of Fig. \ref{f1}(a) indicates the onset $T_c$.
}
\label{f1}
\end{figure}

Figure \ref{f1}(a) shows the pressure dependence of the electrical resistivity as a function of temperature for LaO$_{0.5}$F$_{0.5}$BiS$_2$ (AP). The onset transition temperature $T_c$ is established at $T_{\mathrm{c}}$=2.5 K under ambient pressure. (Inset of Fig. \ref{f1} (a)) By applying pressure up to 0.8 GPa, the $T_c$ increases steeply up to 10.7 K. For pressures above 0.8 GPa, $T_c$ decreases gradually. The measurements indicate a semiconducting behavior in $\rho(T)$, although the resistivity at room temperature gradually decreases with increasing pressure. For higher pressures above 3 GPa, a diamond anvil cell (DAC) is used to measure the electrical resistivity under pressure (Fig. \ref{f1} (b)). An onset transition temperature $T_c$ in the $\rho(T)$ at 2 GPa is observed at 10 K, which is nearly the same onset temperature as at 2.1 GPa in the piston-cylinder cell ($T_c$=10.2 K). $T_c$ gradually changes from 10 K to 5.5 K with application of pressure up to 18 GPa. Furthermore, $\rho(T)$ decreases monotonically with increasing pressure and still exhibits semiconducting behavior above $T_c$ even under the highest pressure of 18 GPa.

\begin{figure}[h!]
\begin{center}
\includegraphics[width=15pc]{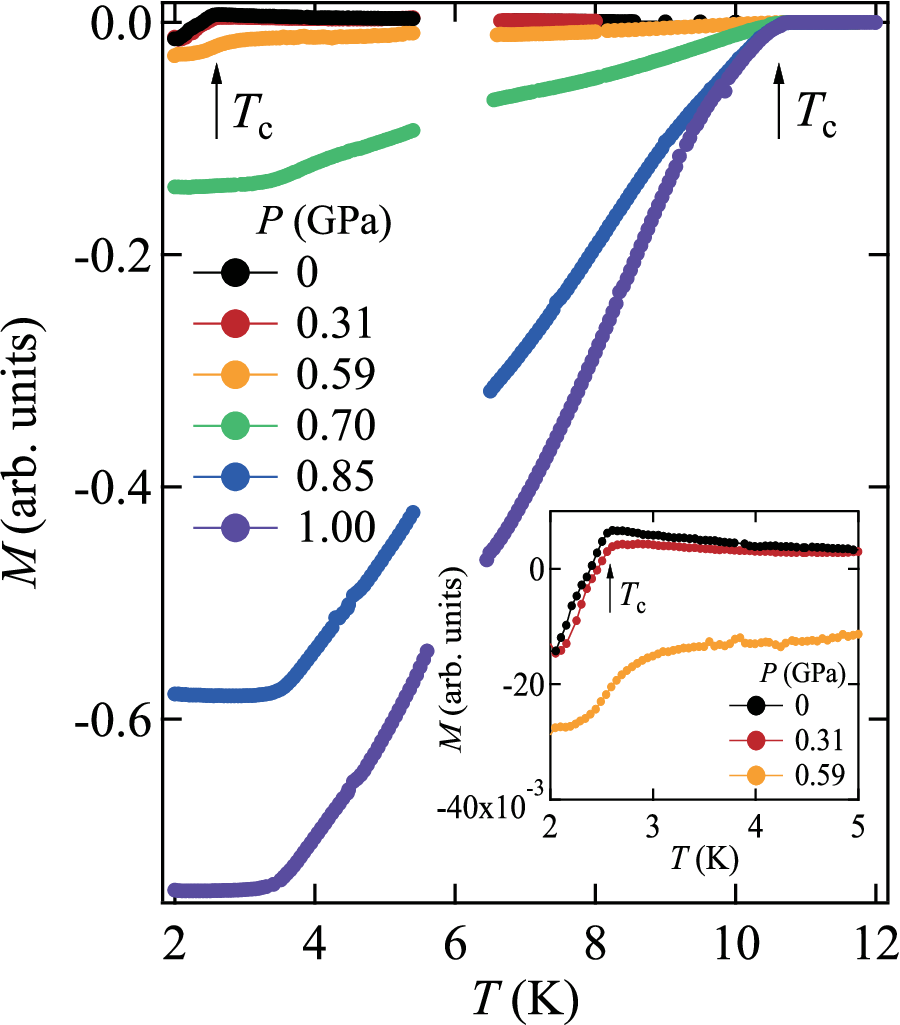}
\end{center}

\caption{Pressure dependence of DC magnetization for LaO$_{0.5}$F$_{0.5}$BiS$_2$ up to 1.0 GPa using a piston-cylinder cell, where a magnetic field of 100 Oe was applied. The arrows indicate the onset superconducting transition $T_c^{M}$. The inset of Fig. 2 shows magnetization near the lower $T_c$=2.5 K.
 }
\label{f2}
\end{figure}

To study $T_c(P)$ up to 1 GPa precisely, magnetization measurements were performed using an MPMS SQUID magnetometer with a piston-cylinder cell. Figure \ref{f2} shows the DC magnetization for LaO$_{0.5}$F$_{0.5}$BiS$_2$ (AP) under various pressures. To estimated the volume fraction under pressure, the magnetization is normalized by a volume fraction for the superconducting transition of lead, which is also the pressure manometer. In Figure \ref{f2}, the magnetization for the superconducting transition of lead, appearing within 6--7 K was erased. The Meissner effect for LaO$_{0.5}$F$_{0.5}$BiS$_2$ (AP) has been confirmed at the onset $T_c$2.6 K at ambient value, which is also the same $T_c$ determined by electrical resistivity measurements (Inset of Fig. \ref{f2}). At 0.59 GPa, the $T_c$ shape becomes broader than the $T_c$ shape at lower pressure. Another $T_c$ appeared near 10.7 K at the same pressure. Above 0.7 GPa, the $T_c$ near 2.6 K completely disappears. The pressure where $T_c$=2.6 K disappears is defined as $P_c$. While the superconducting volume fraction decreases with applied pressure of $P<P_c$ (Inset of Fig. \ref{f2}), it increases when $P>P_c$ (Fig. \ref{f2}). The volume fraction for higher $T_c$ values when $P>P_c$ is 10 times higher than that for the lower $T_c$ values when $P<P_c$. 
The difference of volume fractions of $T_c$ below $P_c$ and above $P_c$ corresponds to that between AP and HP.\textcolor{blue}{\cite{Mizuguchi2012-4}}.

\begin{figure}[h!]
\begin{center}
\includegraphics[width=30pc]{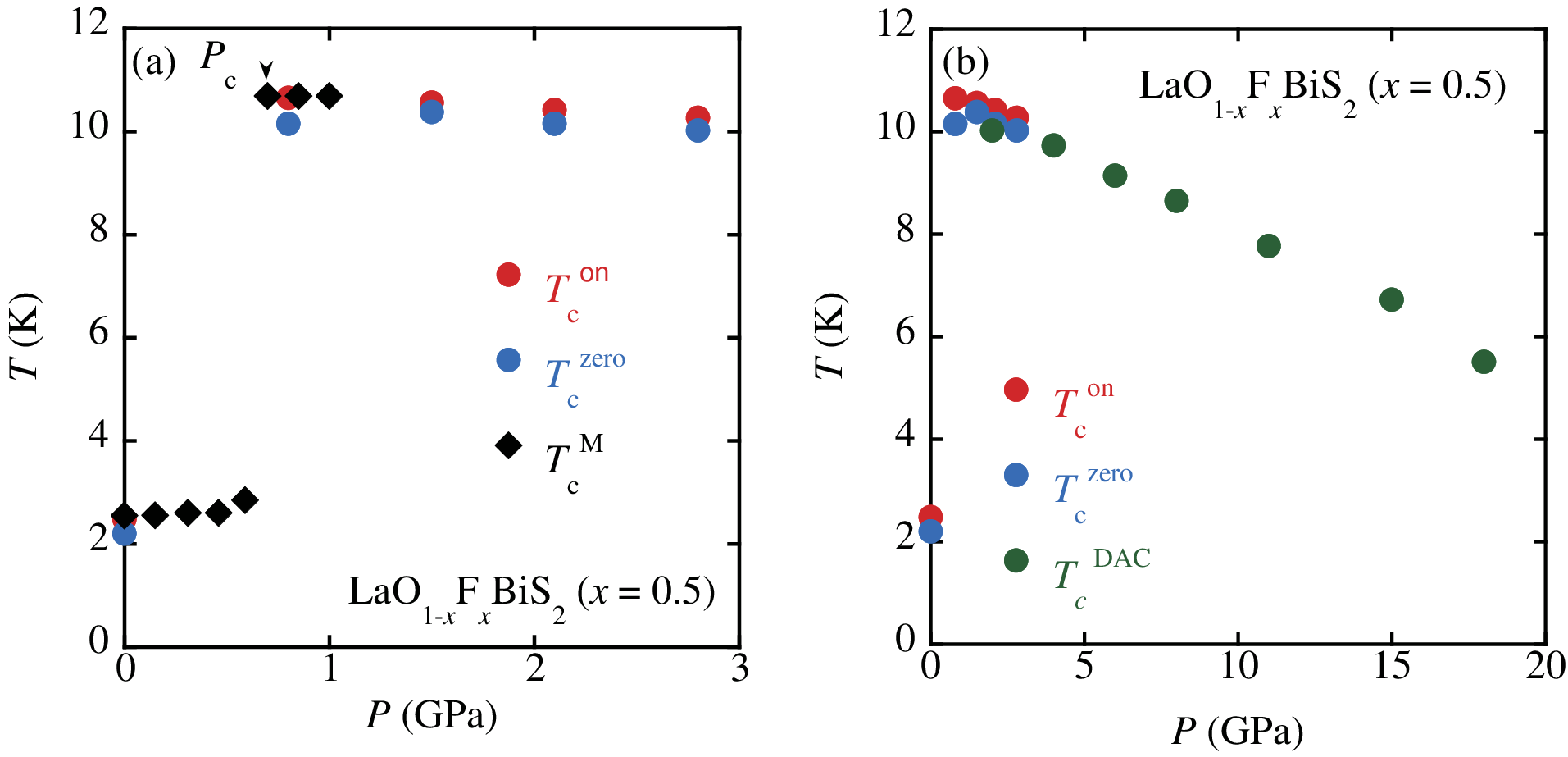}
\end{center}
\caption{
(Color online) 
Temperature versus pressure phase diagrams for LaO$_{0.5}$F$_{0.5}$BiS$_2$, where the superconducting transition was determined by 
the onset temperature $T_c^{\mathrm{on}}$, the zero resistance temperature  $T_c^{\mathrm{zero}}$ in the resistivity measurements, 
and by the onset temperature $T_c^{\mathrm{M}}$ in DC magnetization measurements. 
Fig. \ref{f3}(a) and (b) shows $P-T$ phase diagrams near 1 GPa and up to 18 GPa, respectively.
}
\label{f3}
\end{figure}

The $P$-$T_c$ phase diagrams determined by both the magnetization and the electrical resistivity are shown in Fig. \ref{f3}(a) and (b), where $T_c^{\mathrm{zero}}$, $T_c^{\mathrm{on}}$ ($T_c^{\mathrm{DAC}}$), and $T_c^{\mathrm{M}}$ are defined by zero resistivity measurements, the onset $T_c$ of electrical resistivity in piston-cylinder (in DAC), and the DC magnetization measurements, respectively. In the pressure range of $0\sim0.8$ GPa, $T_c(P)$ is roughly constant ($\sim$ 2.6 K). $T_c$ greatly increases to $T^{\mathrm{max}}$=10.7 K at $P_c$. This maximum is comparable to that for BiS$_2$LaO$_{0.5}$F$_{0.5}$ (HP). \textcolor{blue}{\cite{kotegawa}} Above $P_c$, $T_c$ decreases linearly. The pressure derivative of $T_c$ ($dT_c/dP$) of about -0.28 K/GPa is close to -0.4 K/GPa for the LaO$_{0.5}$F$_{0.5}$BiS$_2$ (HP) sample. 
 
\begin{figure}[h!]
\begin{center}
\includegraphics[width=30pc]{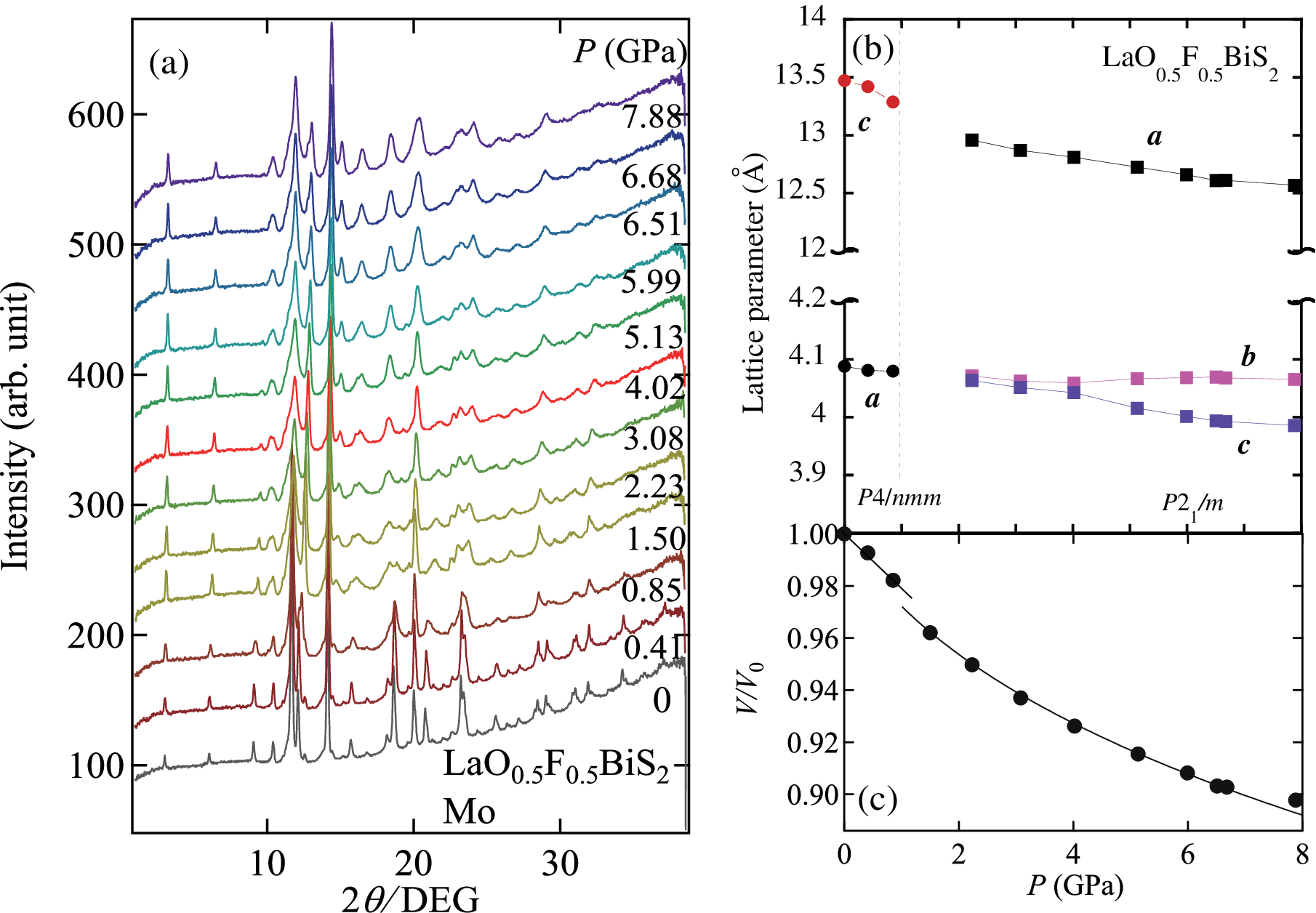}
\end{center}
\caption{
(Color online) 
(a) X-ray diffraction pattern of LaO$_{0.5}$F$_{0.5}$BiS$_2$ using X-rays at room temperature under applied pressure up to 10 GPa. 
(b) The pressure dependence of lattice parameters estimated in both the tetragonal and the monoclinic structure. 
(c) Pressure dependence of the unit cell volume change.
}
\label{f4}
\end{figure}

Figure \ref{f4}(a) shows the X-ray diffraction pattern of LaO$_{0.5}$F$_{0.5}$BiS$_2$ (AP) at room temperature. We assign a tetragonal structure ($P4/nmm$) to the ambient pattern with lattice parameters of $a$=4.0877 and $c$=13.4703 \AA. \textcolor{blue}{\cite{Mizuguchi2012-4}} As shown in Fig. \ref{f4}(a), a splitting near 24$^{\circ}$ appears in the diffraction pattern at 0.85 GPa, which develops with additional applied pressure. Above 1.5 GPa, LaO$_{0.5}$F$_{0.5}$BiS$_2$ undergoes a complete structural phase transition from the tetragonal phase to an another structure.

\begin{figure}[h!]
\begin{center}
\includegraphics[width=30pc]{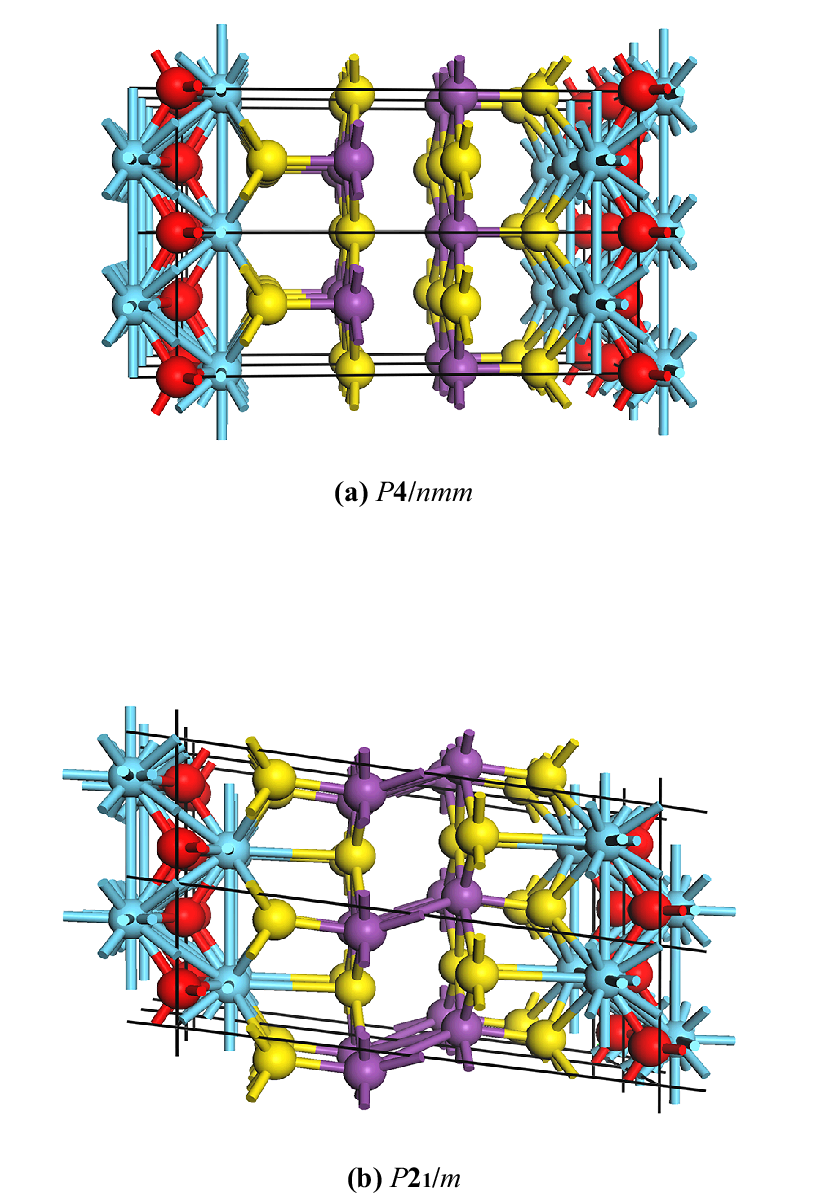}
\end{center}
\caption{
(Color online) 
 2x2x1 unit cells for (a) Tetragonal $P4/nmm$ model at 0.4 GP and (b) monoclinic $P2_1/m$ model at 3.8 GPa viewed from the ac planes.  The Bi-Bi distance at the center of the cell was 3.70 \AA.  

}
\label{f5}
\end{figure}

Analysis of the powder pattern was performed with Accelrys Materials Studio (MS) Reflex software. We confirmed the crystal structure of the AP phase to be a tetragonal $P4/nmm$, as previously reported, which was stable up to about 0.8 GPa. The structure under a pressure of $P \le $ 0.4 GPa is shown in Fig. \ref{f5}(a).  

\begin{figure}[h!]
\begin{center}
\includegraphics[width=30pc]{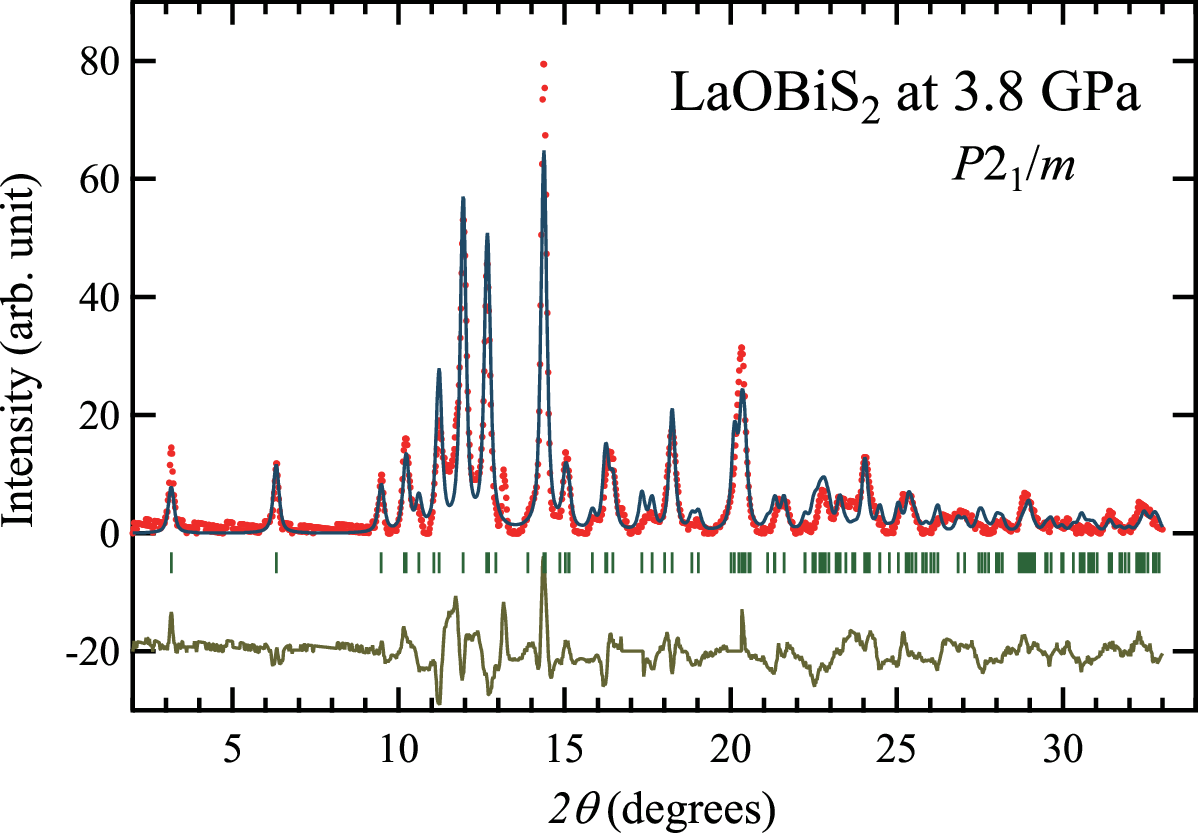}
\end{center}
\caption{
(Color online) 
X-ray diffraction pattern and the Reitveld fitting of LaO$_{0.5}$F$_{0.5}$BiS$_2$ using MoK$\alpha$ X-ray at room temperature under pressure of 3.8 GPa. 
The background has already subtracted.The reliability factor was Rwp=25.5 \%. 
}
\label{f6}
\end{figure}

We performed peak indexing for the HP phase using Accelrys MS X-Cell software. \cite{Neumann2003} These results produced a monoclinic lattice candidate. Rietveld fittings for the HP phase in several experimental runs were not very good, likely because of lattice strain after the phase transition (Fig. \ref{f6}). Therefore, we optimized the atomic positions by using density functional theory (DFT) calculations; specifically, we fixed the lattice constants to the results of Rietveld analysis by Accelrys CASTEP software. \cite{Castep2005} In this analysis, we used the  PBEsol (Perdew--Burke--Ernzerhof for solids) exchange-correlation functional, a generalized gradient approximation (GGA), cite{Perdew2008} with ultrasoft pseudopotentials \cite{Vanderbilt1990}. The energy cut-off for the plane wave basis set was 380.0 eV. The SCF energy tolerance was set to 5$\times$10$^{-7}$ eV per atom. The Monkhorst--Pack grid separation \cite{MP1976} was set to approximately 0.03 \AA $^{-1}$. For geometry optimization, the total energy convergence tolerance was 3.0$\times$10$^{-6}$ eV per atom, with a maximum force tolerance of 0.007 eV/\AA , a maximum displacement of 3.0$\times$10$^{-4}$ \AA , and a maximum stress tolerance of 0.01 GPa. Using these parameters, we obtained a monoclinic $P2_1/m$ model, as shown in Fig. \ref{f4} (b). The powder patterns we obtained for both Non-F doping ($x$=0) and F-doping ($x$=0.5) samples could be explained by this model. The estimated lattice parameters for the tetragonal and monoclinic structures are shown in Fig. \ref{f4}(b).
%%Examples for lattice parameters and atomic coordinates are summarized in Tables 1 and 2. 
The pressure dependence of lattice parameters of the AP phase and HP phase obtained by Reitveld analysis are shown in Fig. \ref{f6}. The lattice angle $\beta$ initially increases with applied pressure and then becomes constant around 97$^{\circ}$ $\sim$ 98$^{\circ}$.

It is not well understood whether the poor fit of the simulated diffraction pattern in the HP phase for both $x$=0 and 0.5 is caused by alignment under hydrostatic pressure or by an essential quality of the phase transition. The diffraction pattern fit for the HP phase could not modified, even when we attempted to use He gas instead of Daphne as the pressure medium. Additionally, the diffraction pattern remains unchanged even when measured after annealing at 150 $^{\circ}$C for 2 h under a pressure of 5 GPa. The pattern may only changes when the temperature is increased to around 1000 $^{\circ}$C using laser heating. Moreover, the diffraction pattern recovers to that of the AP phase; the HP phase is not quenched at ambient pressure.

We should now consider the characterization of the HP phase and the mechanism of the phase transition from AP to HP using the crystal structures obtained by the optimized DFT calculations. The phase transition from AP to HP is caused by sliding between the two BiS$_2$-layers along the $a$-axis, causing the lattice structure to become monoclinic under pressure. This mechanism was reproduced by the molecular dynamics simulations of DFT calculations at 300 K under 10 GPa. Moreover, the structure under the higher pressure can be calculated using DFT by including the experimentally obtained lattice parameters. In the data for the HP phase (5.6 GPa) for the $x$=0.5 sample, the Bi--Bi distance between BiS$_2$ layers is approximately 3.52 \AA. These results do not clearly show that the Bi--Bi bond exists at this pressure; however, because this Bi--Bi distance is less than 4.14 \AA (twice the van der Waals radius of bismuth), the Bi--Bi interaction would exist. This Bi--Bi distance shrinks to 3.37 \AA \ under a pressure of 10.0 GPa. We estimate the distance to be 3.19 \AA \ under 20 GPa and 3.09 \AA \ under 30 GPa; the distance continues to shrink with additional applied pressure. The distance under a pressure of 20 GPa reaches 3.12 \AA, two times larger than the atomic radius of Bi; this result clearly shows that a Bi--Bi bond forms, causing the coordination of Bi to become 7.

From these Rietveld analysis, the pressure-generated diffraction profile fits best to a monoclinic phase with a space group $P$2$_1/m$. The pressure-induced phase transformation is generated at 0.8$\sim$1.5 GPa, which corresponds to  $P_c$.  The lattice parameters $b$ and $c$ for monoclinic structures gradually separate with applied pressures up to 8 GPa. Additionally, a small volume change occurs at $P_c$ as shown in Fig. \ref{f4}(c). In the pressure region $P<P_c$ for the tetragonal phase, the isothermal compressibility of the unit cell volume can be calculated by a first pressure derivative of unit cell volume, which is estimated as $-d(V/V_{0})/dP$=0.0089 GPa$^{-1}$ (bulk modulus 112 GPa), where  $V_0$ is the unit cell volume at ambient pressure. In the pressure region $P>P_c$ for the monoclinic phase, the unit cell volume gradually decreases with application of pressure up to 10 GPa. This structural phase transition is also induced in the LaOBiS$_2$ sample, which will be expanded upon in a later report.\textcolor{blue}{\cite{tomita}}
Therefore, we conclude that the structure at higher pressures ($P>$0.7 GPa) with $T_c$=10.7 K is monoclinic. 
Another important point is that LaO$_{0.5}$F$_{0.5}$BiS$_2$(AP) is superconductive for both the tetragonal structure below $P_c$ and the monoclinic structure above $P_c$. Thus, the phase transition at $P_c$ indicates a pressure-induced superconducting-superconducting transition similar to transitions in CaC$_6$ ($P_c$=10 GPa) and Bi ($P_c$= 8 GPa).\cite{debessai, Ilina}
This structural change gives an important effect on the superconducting BiS$_2$ layer to cause a $T_c$ rise.
As described in reports on FeAs-layered materials\textcolor{blue}{\cite{lee}}, the bond angle of As-Fe-As in the superconducting FeAs layer influences $T_c$. This BiS$_2$-layered material may also have an important structure parameter that determines $T_c$, similar to the bond angle in FeAs-layered materials. If we can control this structural parameter for BiS$_2$-layered materials by using chemical doping and external pressure, further enhancement of $T_c$ is expected.

\section{\label{sec:SUMMARY}Summary}
We performed electrical resistivity, magnetization, and X-ray diffraction measurements on a new BiS$_2$-layered compound LaO$_{0.5}$F$_{0.5}$BiS$_2$ (AP) under pressure to determine the $P$-$T_c$ phase diagram up to 18 GPa. $T_c$ greatly increases from 2.6 K to 10.7 K at an applied pressure of 0.7 GPa, where a structural phase transition from a tetragonal ($P$4$/nmm$) to a monoclinic ($P$2$_1/m$) structure occurs. Additionally, a $T_c$ value of 10.7 K in the high-pressure regime ($P_c>$0.7 GPa) appears in the monoclinic structure. 
The maximum $T_c$10.7 K for AP processing with the monoclinic structure is comparable to its $T_c$10.6 K for HP processing under ambient pressure and high pressure \textcolor{blue}{\cite{kotegawa,Mizuguchi2012}}. Therefore, the HP-processed LaO$_{0.5}$F$_{0.5}$BiS$_2$ (HP) may contain the monoclinic phase.

\begin{acknowledgments}
This work was partially supported by a Grant-in-Aid for Young Scientists (B) (No. 19740208) and the Strategic Research Base Development Program for Private Universities (2009, S0901022), and was subsidized by the Ministry of Education, Culture, Sports, Science and Technology (MEXT) of Japan.
\end{acknowledgments}

\newpage
%%%%%%%%%%%%%%%%%%%%   BiS2LaOF,, figure 1  %%%%%%%%%%%%%%%

%%%%%%%%%%%%%%%%%%%%%%%%%%%%%%%%%%%%%%%%%%%%%%%%%%%%%%%%%%%

%%%%%%%%%%%%%%%%%%%%   BiS2LaOF,, figure 2  %%%%%%%%%%%%%%%
%%%%%%%%%%%%%%%%%%%%%%%%%%%%%%%%%%%%%%%%%%%%%%%%%%%%%%%%%%% 

%%%%%%%%%%%%%%%%%%%%   BiS2LaOF, figure 3  %%%%%%%%%%%%%%%

%%%%%%%%%%%%%%%%%%%%%%%%%%%%%%%%%%%%%%%%%%%%%%%%%%%%%%%%%%%

%%%%%%%%%%%%%%%%%%%%   BiS2LaOF,, figure 4  %%%%%%%%%%%%%%%

%%%%%%%%%%%%%%%%%%%%%%%%%%%%%%%%%%%%%%%%%%%%%%%%%%%%%%%%%%%

%%%%%%%%%%%%%%%%%%%%   BiS2LaOF,, figure 5  %%%%%%%%%%%%%%%

%%%%%%%%%%%%%%%%%%%%%%%%%%%%%%%%%%%%%%%%%%%%%%%%%%%%%%%%%%%

%%%%%%%%%%%%%%%%%%%%   BiS2LaOF,, figure 6 %%%%%%%%%%%%%%%

%%%%%%%%%%%%%%%%%%%%%%%%%%%%%%%%%%%%%%%%%%%%%%%%%%%%%%%%%%%

\end{document}